\documentclass[a4paper,12pt]{article}%
\usepackage{amsmath}%
\usepackage{amsfonts}%
\usepackage{amssymb}%
\usepackage{graphicx}
\begin{document}

\title{Beyond Markov Chains, Towards Adaptive Memristor Network-based Music Generation}
\author{Ella Gale\thanks{Corrosponding author}, Oliver Matthews,\\ Ben de Lacy Costello and Andrew Adamatzky
\\The University of the West of England}

\maketitle

\begin{abstract}
We undertook a study of the use of a memristor network for music generation, making use of the memristor's memory to go beyond the Markov hypothesis. Seed transition matrices are created and populated using memristor equations, and which are shown to generate musical melodies and change in style over time as a result of feedback into the transition matrix. The spiking properties of simple memristor networks are demonstrated and discussed with reference to applications of music making. The limitations of simulating composing memristor networks in von Neumann hardware is discussed and a hardware solution based on physical memristor properties is presented.
\end{abstract}

\section{Introduction}

The human universals~\cite{M4} are traits found in all human cultures since the Upper Paleolithic and are unique to humanity; music is on this list of around 370 concepts and behaviours including examples like: dance, hope, language, fire, fear of death, cooking, prohibition of murder, hairstyles and other behaviours both similarly dramatic and banal. Music's role in human culture is related to sexual attraction, social cohesion, relaxation and communication (see ~\cite{M5} for a recent review from the anthropological context). It is believed that, like some other human universals, music may be a product of the structure of the mind~\cite{M4}, and thus a by-product of human evolution. However, the popular idea that music is a universal language or pre-language has been resoundly disproved, as far back as 1940~\cite{M6}, by cross-cultural studies that showed that the emotional resonance of music is a culturally learned response. 

The combination of two human universals, anthropomorphisation and tools, would suggest that the best tool would be human-like and thus it's not surprising that, after the invention of computers, artificial intelligence, A.I., (namely the desire to create an intelligent machine) would be an area of active research. A.I. has had some successes such as learning-classifying systems and neural networks, however the creation of creative A.I.s has had fewer successes, and the harder problem of creating a self-aware and conscious machine intelligence has suffered from even less progress.

Music generation is a good problem to tackle if one is interested in making a creative A.I., furthermore, if music does arise as a result of brain structure, then it might be fruitful to approach the problems of neuromorphic engineering (that of making brain-like computers) by creating a composing brain utilising a human-selection process on the output music: it's easier to recognise a melody than brain-like activity in a neural network. This approach will have the added drawback (or perhaps benefit) of adding a cultural bias to the music.

The study of creativity is a large and multidisciplinary area, with competing definitions and a lack of consensus, however, a recent attempt at formalising creativity and describing the actions of creative A.I. agents requires two modes of learning: A. an adaptive predictor or model of the growing data history and B. a reinforcement learner~\cite{M3}. The agent learns about the world around it, compresses and stores that data and as such makes predictions about the future, this model encodes the known structure of a certain style of music, thus making the output similar to music in the same style, but by using reinforcement learning tuned to novelty, the new music is different enough to be interesting.

Markov chains have been well exploited in music generation~\cite{M7}. The Markov Hypothesis states that a future state of a sequence depends only on the last state. Usually a matrix of note transitions is seeded with a corpus of music of a particular style and music is generated via a random walk. While often effective, it has a number of drawbacks; the biggest  is that music has an underlying structure and requires long-term order which contravenes the Markov Hypothesis~\cite{M8}.

Unconventional computing is a branch of computing that aims to go beyond the von Neumann models of computation and includes, but is not limited to, methods like chemical computation, biological computation, cellular automata, quantum computing, neuronal computing~\cite{M9}. Various unconventional computing approaches have been applied to music generation, such as cellular automata music generation, sonifying \textit{Physarum polycephalum} and sound synthesis using a neuronal network (wetware). Using memristors is considered a unconventional computing approach due to their novel communication interactions~\cite{243} and similarity to the neurons~\cite{247,248}.

Memristors are the recently discovered~\cite{15} 4$^\mathrm{th}$ fundamental circuit element~\cite{14}. The memristor changes its resistance as a function of the amount of charge that has passed through it (which is also proportional to voltage). Unlike the other three fundamental circuit elements (the resistor, capacitor and inductor) the memristor is non-linear and possesses a memory. Memristors have also been compared to neurons in the brain due to their spiking response to changes in input. 

For the auto-generation of music, we are interested in four properties of memristors: their non-linearity, their time-dependence, their memory and their spiking response. As a network of memristors would necessarily possess a memory that goes beyond the previous state \footnote{technically the memristor's memory is dependent on its entire history from $-\infty$ to now, in practice it is possible to `zero' a memristor's memory}, music generation using a memristor networks offers us a route to go beyond Markov chaining.

Memristors have been used as synapse analogues in STDP (Spiking Time Dependent Plasticity) neural networks~\cite{214}. Here, we plan to use memristors as the connections between a graph of musical notes, where the memristors can modify their connection weight non-linearly with the number of times one musical note follows on from another in a piece of seed music. This will create a weigthed graph, which can be built in the lab by connecting memristors. In this paper, we shall simulate such a graph by simulating the memristor connections, as in~\cite{A1}. 

A network of memristors can spike, and these spikes are believed to be deterministic and related to the change in voltage across a memristor. These voltage changes and spikes can propagate across a network through time in a complex manner (as the voltage change from one spike will cause a voltage change in memristors further along the network, causing further spikes and so on). Thus, the spike interactions can be used to `play' music by choosing which notes follow on from one another (in a Markov chain approach). 

There are two timescales that interact in a memristor network and which can give rise to altering tempo of played notes. The first is the relaxation time of a memristor after its spike (this is related to the memristors memory). The second is related to the length of the wires between the memristor and the time taken for a signal to propagate from one spiking memristor to the rest of the network. 

These three aspects: the seeded network, the spikes and the time dependence can interact to give complex behaviour. However, each spike will alter the structure of the network, allowing the system to change over time, leading to developing new patterns in its musical style to `evolve' (or possibly allowing it to get stuck in a stable attractor).

In this paper, we will discuss the methodology and challenges for building such a network, by simulating a demo network, demonstrating the spikes responses in a simple real memristor network, simulating a simplified version of the spiking seeded network and finally discussing whether such a spiking seeded network can be fully simulated in a computationally tractable way using standard von Neumann architecture.

\section{Building the Memristor Network}

\subsection{Setting up the graph}

\begin{figure}
	\centering
		\includegraphics{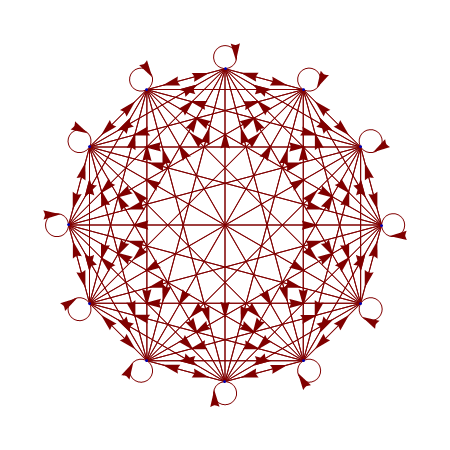}
	\caption{A self-linked, complete directed k-graph. Note that the forwards and backwards connections are drawn to overlap here.}
	\label{fig:12kgraph}
\end{figure}

\begin{figure}
	\centering
		\includegraphics{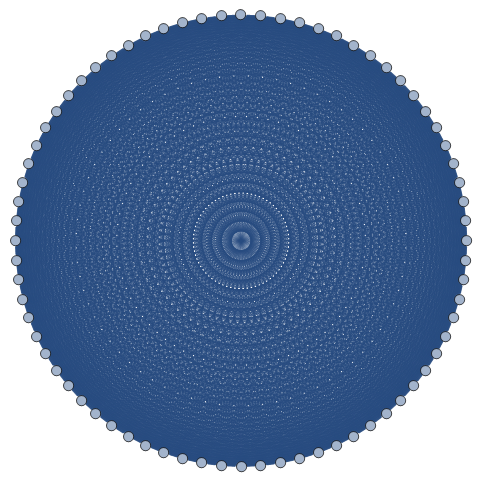}
	\caption{Complete k-graph connected for the 6 octaves of a piano. The self-connections have not been drawn for clarity.}
	\label{fig:Piano}
\end{figure}

For this work we will consider a musical range of only two octaves, stretching from C4 to B$\flat$5, for a total of 24 notes. As any note could potentially follow on from any other, the graph of all possible links would be a reflexive directed k-graph of 24 nodes and 576 vertices. We show, as an example, a fully connected directed k-graph for 12 notes (an octave) in figure~\ref{fig:12kgraph}. For comparison, a network for a full standard piano would require 88 nodes, requiring 7744 memristors to model, as shown in figure~\ref{fig:Piano}. In the real network, a transition (e.g. A5$\rightarrow$B$\flat$5) would be recorded by an ammeter in series with the memristor). 

Similarly, the timing of the notes was also constructed by a network. In the actual device, we would expect the memristor spikes to provide the tempo information, for out simplified model a second (much simpler) network built for the tempo analogously as for the notes in the melody. The tempo was broken down into 9 components: semiquaver, quaver, crochet, minim, their dotted versions and the breve.

The memristor network would be held at a constant voltage and as the memristor spikes, these spikes then propagate around the network, each spiking memristor is transiently the source of the $\Delta$V perturbation and each other memristor is the drain. In our simulation, this will be modelled by moving the source to the node associated with the previously played note at each step.

\subsection{Memristive Connections}

\begin{figure}
	\label{fig:GVsTMusic}
	\centering
		\includegraphics[scale=0.75]{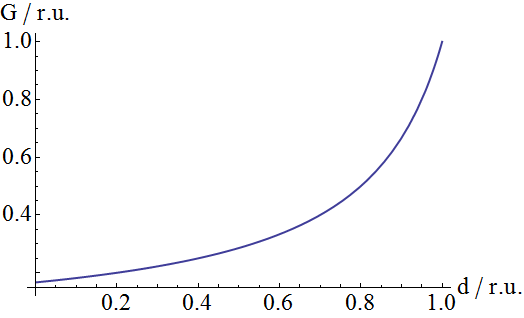}
	\caption{Conductance profile for a continuously charged memristor. This is also how the connection weight for a memristor-based connection.}
\end{figure}

The `probability' of a transition between two notes occurring will be related to the connection weight of that vertex. In the memristor network, the connection weight is the conductivity of the memristor. Figure~\ref{fig:GVsTMusic} shows the conductivity of am memristive connection increasing under a constant voltage (with a linear conductivity profile shown for comparison). As each transition is either heard or created, the memristor conductance moves up this curve. As music is a directed graph (it matters whether we go from C$\rightarrow$A or A$\rightarrow$C) there will be a second memristor, wired up the other way round, which goes down the curve. 
This property means that the reverse transition is less likely if the melody has just performed the transition. 

Memristance is defined~\cite{14} as:
$$
V = M(t) I \: ,
$$
we will use the Mem-Con model of memristance~\cite{259}
$$
M(t) = M_{e}(t) + R_{\mathrm{con}(t)}
$$
which has the advantage of having been fit to the devices in our lab, so we can later use measured experimental values as in~\cite{255}. In this paper, we use reduced units, i.e. the conductance is measured in terms of device properties such as the size and resistivity of the material. The conductivity of a memristor, $G(t)$, is given by
$$
G(t) = \frac{1}{M(t)} \: ,
$$
and as the connection weight in the graph is simply the conductivity, it is also represented by $G(t)$.

\subsection{Seeded graphs}

To seed the melody network, we converted several different pieces of musical melody to a list of notes in the key of C Major, transposing to the key of C and mapping them to the two octave range we have available. To seed the tempo network, the tempo was converted to reduced units where a time of `1' is equal to 1 crochet, this allowed us to normalise for beats-per-minute variations between the seed music. This approach divorces the tempo from the notes, which  we felt was accurate as melody rarely correlates the speed of the note to its pitch (however as many a base singer will say, the backing baseline of harmonising choral pieces is usually the beat and thus includes less variations in the tempo).

From a frequency analysis of the number of times a transition happened, a transition matrix was populated with the expected conductance values as based on the memristor conductivity curve in figure~\ref{fig:GVsTMusic}. Different musical seeds created networks with different structures. We chose to investigate three distinct styles of musical melody, namely  jazz standards, rock'n'roll as exemplified by Elvis (his faster tracks) and light opera as exemplified by Gilbert and Sullivan. The three Jazz standards were: `How high the moon',`Ain't that a kick in the head', `I've got the world on a string'. The Elvis tunes were: `All shook up', `Burning love' and `Jailhouse rock'. The chosen Gilbert and Sullivan classics were three solos taken from Pirates of Penzance: `Modern Major General', `When a felon' and `Better far to live and die'. Specifically, the primary vocal line was taken for each as the melody.

\begin{figure}
	\centering
		\includegraphics[scale=0.25]{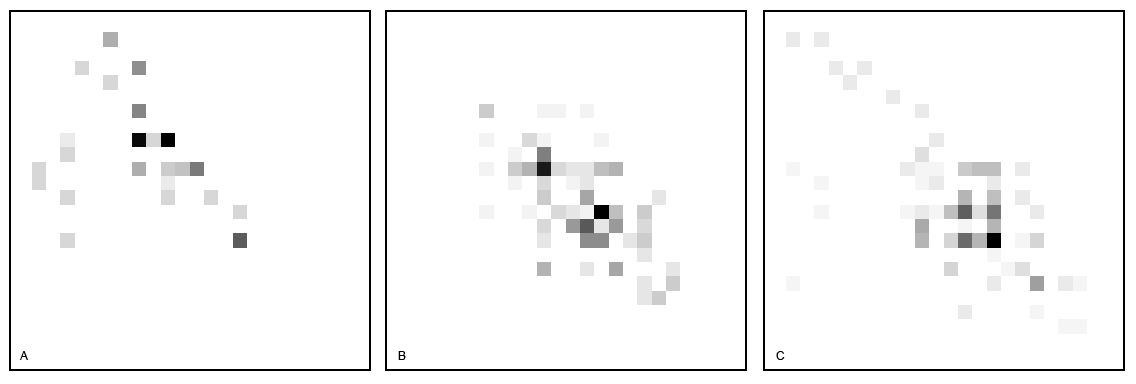}
	\caption{Example connection matrices for memristor networks seeded with the melody line of jazz standards: A, How high the moon, B, One for my baby, C Ain't that a kick to the head. The darker the colour, the more often that transition is heard.}
	\label{fig:ComboMusic}
\end{figure}

For example fig~\ref{fig:ComboMusic} shows example graphs for the melody lines for the three jazz standards, specifically. The graphs tend to be sparse as a single melody is repetitive and does not cover a huge note-space. The largest connection weights tend to be on or close to the diagonal due to the fact that repetition of a note is common when singing a phrase, and because the further from the diagonal the larger the jump between the notes and the human voice has difficulty with larger jumps.

\section{Music Generation}

Having seeded our network, we will now discuss how to use it to create music. We shall start by examining how memristor networks act, and then consider the approximations that must be made to model them in simulation. 

\subsection{Memristor Networks}

\begin{figure}
	\centering
		\includegraphics[scale=0.75]{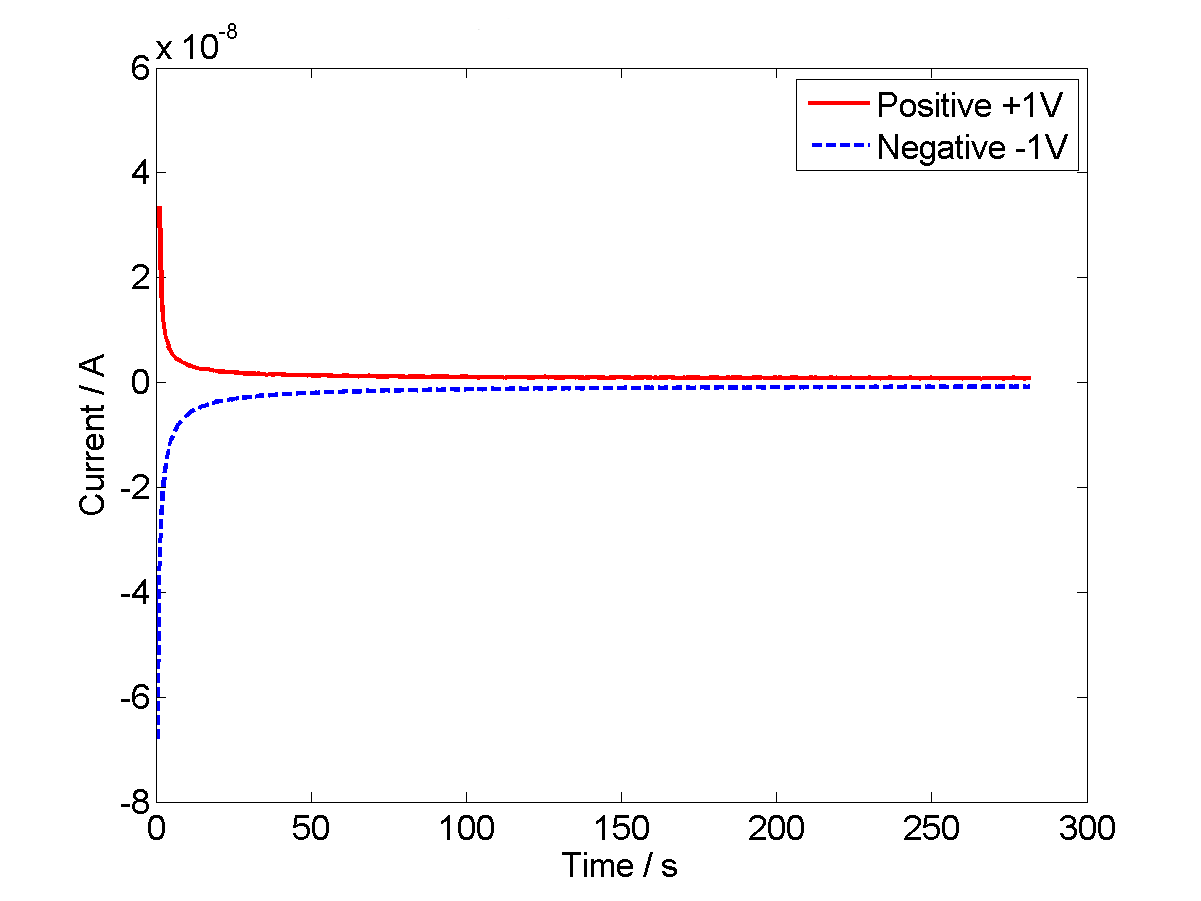}
	\caption{Example positive and negative spikes.}
	\label{fig:SingleSpike}
\end{figure}

Fig~\ref{fig:SingleSpike} shows an example of memristor spikes. There are a recovery time when the system relaxes to it's long term value. This takes around [20]s and may be tunable by varying device parameters. Generally, the larger $\Delta V$ the larger the spike.

\begin{figure}
	\centering
		\includegraphics[scale=0.75]{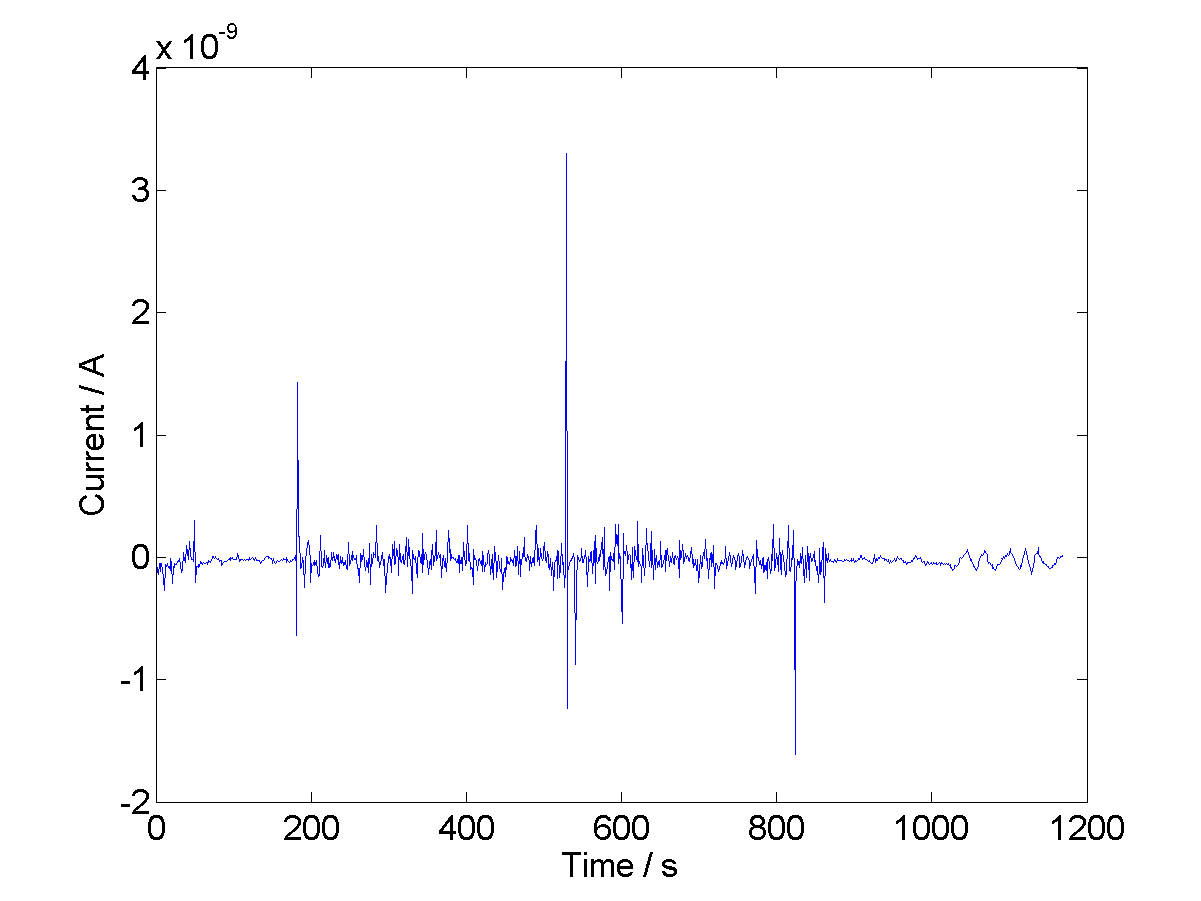}
	\caption{An example of spike patterns through a very simple memristor network}
	\label{fig:2MemristorSpike}
\end{figure}

\begin{figure}
	\centering
		\includegraphics{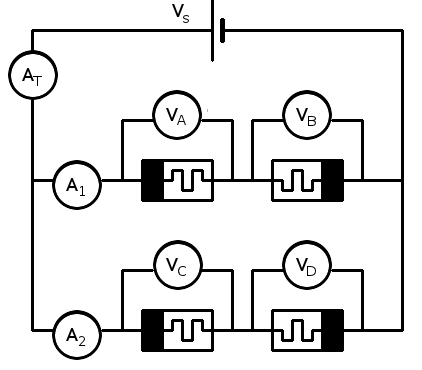}
	\caption{A scheme for two note transitions as an example.}
	\label{fig:CircuitScheme}
\end{figure}

\begin{figure}
	\centering
		\includegraphics{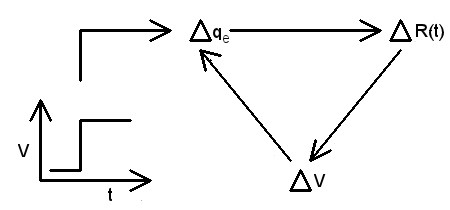}
	\caption{The process of spike creation and continuation.}
	\label{fig:Process}
\end{figure}

Intriguingly, these spikes give rise to complex behaviour. Consider the circuit in figure~\ref{fig:CircuitScheme} A d.c. voltage source is applied from a Keithley 2400 sourcemeter (drawn as a battery, as is standard in such circuits), for a single memristor this would apply a sharp step function from 0V to the set voltage at the first step and then hold it there. Figure~\ref{fig:SingleSpike} shows the response of a memristor to such a voltage. When there are multiple memristors in a circuit, the spike from the voltage application isn't there, instead you get complex behavior like that seen in figure~\ref{fig:2MemristorSpike}, this is response of ammeter A$_\mathrm{T}$. We suspect that this is due to the sudden voltage step occurring at slightly different times across the 3 memristors. Each current spike causes a change in resistance across the memristor it reaches, which causes a change in voltage $\Delta V$ which then causes another current spike, this can bounce around the network indefinitely if provided with an energy source (namely the applied constant voltage). There are two routes by which the memristors can interact, the first is the creation and movement of current spikes, $\Delta q_{e^{-}}$, as in extra charge, $q$, is drawn from the source, and this alters the resistance. The second is a change in resistance, $\delta R$ which causes a change in voltage (as $\frac{1}{\frac{1}{V_{A}+V_{B}}+\frac{1}{V_{C}}}$, which itself causes a current spike. Figure~\ref{fig:Process} summarises this interrelation.

\subsection{Time dependence}

The time dependence due to the delay in signal (current) propagation in a memristor network has been discussed. The other relevent time is the relaxation time, $\tau_r$. When a memristor representing a transition from $X\rightarrow Y$, where $X,Y \in \{C D\flat D E\flat E  F G\flat G A\flat A B\flat B \}$, spikes and alters its resistance, the reverse transition, that of $Y\rightarrow X$ is slightly inhibited because both memristors between nodes X and Y have been altered. The lifetime $\tau_r$ defines how long the memristors take to equilibrate. Before that point, the spikes are smaller if in the same direction and larger if in the opposite way.

These interactions in timing will also cause the spikes to occur at a non-regular rhythm, avoiding boring musical tempo. However, the interactions that give rise to the oscillations lead to concept of a beat. 

\section{Modelling a spiking memristor network for music creation}

If we have a seeded memristor network and turn on a constant voltage, we should create spiking network. If each spike is taken as making a transition from one note to another, then the network's activity will generate music. Also, as the network generates music, it is also learning, so the network will respond to the music it creates and alter the connectivity of the k-graph and thus the music created by the network. 

\subsection{Problems with attempting to model a memristor network}

There are two main problems with modelling such a network. The first is the problem of modelling transient $\delta V$. This is relatively easy to solve. In the real network we would set the wired network up and record what it produces. In the simulation, we don't need a background voltage to power the simulation and can thus set the current note as the source (and itself and all others as a possible drain). 

The second problem is more intractable. We need to know when and if a memristor will spike. It is not known if the spikes are probabilistic or deterministic in nature; this is a current area of investigation. To model to the system as deterministic, we would need to descretise time very finely and model the state of everything in the network at once. If the system is chaotic or near the edge of chaos (a very real possibility) any approximations or course-graining of the system will result in an extremely inaccurate simulation. Furthermore, it is not currently known what causes the spikes, we suspect that those measured in fig~\ref{fig:2MemristorSpike} are the addition of spikes from the individual memristors. But we don't know what causes these individual memristors to spike or not spike. Also, the network as drawn in part one of this paper, is not a standard electronic circuit that can be entirely resolved into series and parallel relationships, making the network as a whole non-simple and difficult to model. 

Finally, for a one octave memristor network we require 144 memristors, to do an entire piano's range we would need 7,744 (plus 81 for the tempo). We can not necessarily make the approximation of considering one memristors against a `mean-field' background of the other memristors, due to the almost instantaneous~\footnote{As energy can not be created or destroyed, a change in voltage should change the voltage drop across the rest of the network instantaneously. Whether this change is actually instantaneous or proceeds at the speed of light is a question for relativity physicists} $\Delta V$, thus even with our simplified version we have a 576-body problem to attempt to calculate. We suggest that solving such a problem with standard von Neumann computer architecture will be computationally intractable (although it is not theoretically impossible). 

Our obvious solution to these issues is to build a non-von Neumann computer architecture, i.e. to actually build a network of 576 memristors. However, before undertaking such an endeavour, it's worth doing a preliminary, highly simplified simulation to check that a seeded memristor network would be of use to generating music and to see if the non-linearity of the memristor model by itself does not offer interesting aspects to procedural music composition. 

\subsection{Using a simplified memristor network model to perform non-Markovian music generation on a pre-seeded network}

As described in the previous section, the `roving' $\Delta V$ will be modelled as $V$ against a background of $0V$, in that each note will be set to the voltage source in turn. This is a gross simplification of the laws of physics but will serve to course grain the effects of a network. The drain will be connected to the drain of all other nodes (including the drain of that node, which allows for a self$\rightarrow$self transition. 

There is a function $p(t)$ which controls whether a memristor will spike or not. We suspect that the memristor network is deterministic and chaotic in form, but have no current knowledge of this function. Therefore, we shall take the simplification of assuming that the pseudo-random chaos can be coarsely representated as the pseudorandom values from a Gaussian random number generator. Thus, we will talk about the probabilities of a transition between $X$ and $Y$ occurring, $p_{(X\rightarrow Y)}$ as being a product of the connection weight and our unknown function $p(t)$, which is itself set as a pseudorandom number, $p(x)$. Thus, 
$$
p_{(X \rightarrow Y)} = G_{(X \rightarrow Y)} p(x) \; .
$$

And the next note, $n+1$, is determined by the maximum of this product over the set of all a possibilities, i.e.:
$$
p_{(n \rightarrow +n)} = \mathrm{Max}[ \{ p_{(X \rightarrow Y)} : X, Y \in \{ C4:B\flat5 \} \{ p(x)_{T} \} ] \:,
$$
where $T$ is the set of all 576 possible transitions.

After a given connection has fired, we slightly increase it's state along the memristive curve (to reflect the current that flowed through it as part of the spike) and use this new state for calculating future transitions. To model the relaxation time, the memristor that has spiked on step $n$ is artificially moved down the memristor curve to a quarter of it's value for step $n+1$ and half for step $n+2$, this substantially reduces the likeliness of it firing again until step $n+3$ where it is set to its new (increased) weight. To model the reverse note connections (and prevent the over-occurance of the odd musical structure of $X \rightarrow Y \rightarrow X \rightarrow Y \rightarrow X ... \infty, {X,Y} \in \{ C4:B\flat5 \}$, the reverse transitions are decremented rather than incremented at step $n$ and similarly reduced to a quarter and half their new value on steps $n+1$ and $n+2$. The music is started on note $C4$ and the first note of the tune taken from the first transition from that note. 100 notes were generated for each tune.

\begin{figure}
	\centering
		\includegraphics{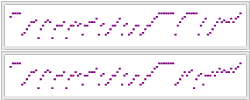}
	\caption{Generated music based on the 'How high the moon' connectivity matrix. Top is without feedback into the connectivity matrix, bottom included feedback, both have the same pseudorandom number input. The connectivity matrix is altered and the alteration is slow. }
	\label{fig:MoonCombo}
\end{figure}

\begin{figure}
	\centering
		\includegraphics{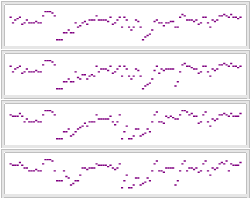}
	\caption{The effect of feedback into the transition matrix. Here the matrix used was the combination which was seeded by all three jazz standards. In order, they are melodies generated by the original triple matrix, and those generated after 1,000,10,000 and 100,000 notes generated respectively.}
	\label{fig:TripleEvolved}
\end{figure}

Figure~\ref{fig:MoonCombo} shows an example of generated music from the connectivity matrix seeded with 'How high the moon'. Despite the simplicity, it sounds like music rather than random notes. The top subfigure shows music output from a static matrix, the bottom shows the effect of allowing the transition matrix to be seeded by the musics it's generating, and thus can only be seen at the end. This is what we want, as we don't want the music generator to change too quickly. 

Figure~\ref{fig:TripleEvolved} shows further examples of the plasticity of the transition matrix. This figure clearly shows the strong effect of the transition matrix on the music generated, but we can see that over time and repeated generated notes, the melody is changing. 

\subsection{Results from our modelled network}

\begin{figure}
	\centering
		\includegraphics[scale=0.35]{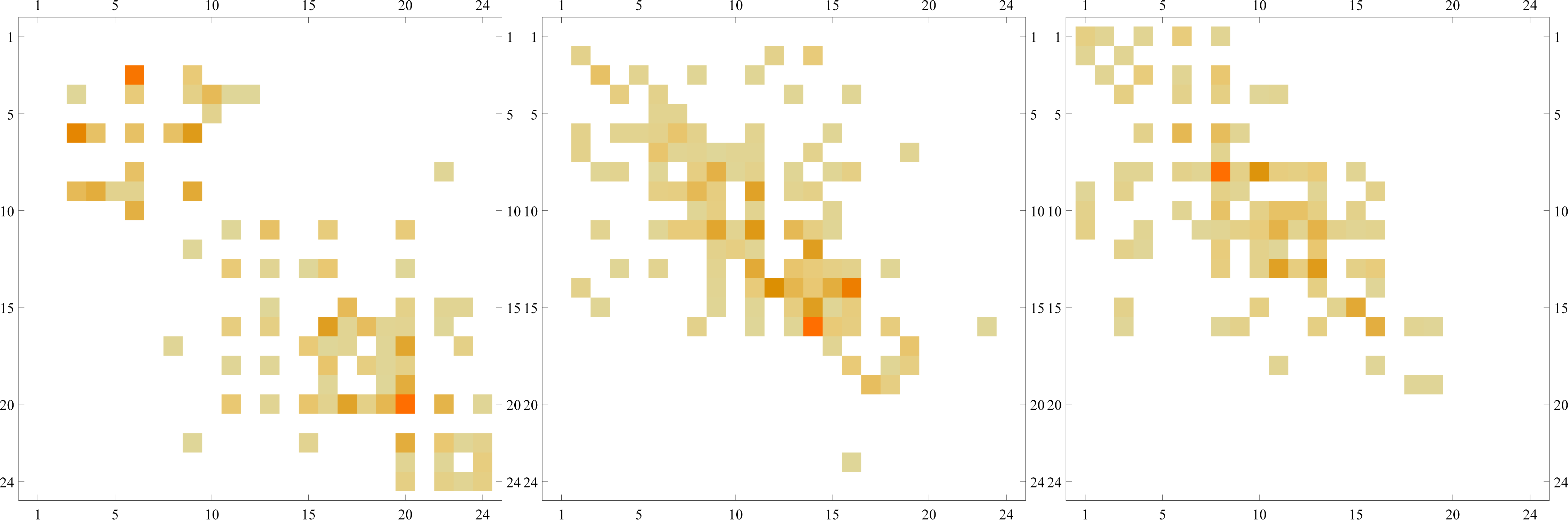}
	\caption{The connection matrices as seeded from three pieces of music in each genre: A. is Jazz standards, B is the Elvis rock'n'roll and C is the Gilbert and Sullivan light opera. }
	\label{fig:ComboMusicNote}
\end{figure}

\begin{figure}
	\centering
		\includegraphics[scale=0.35]{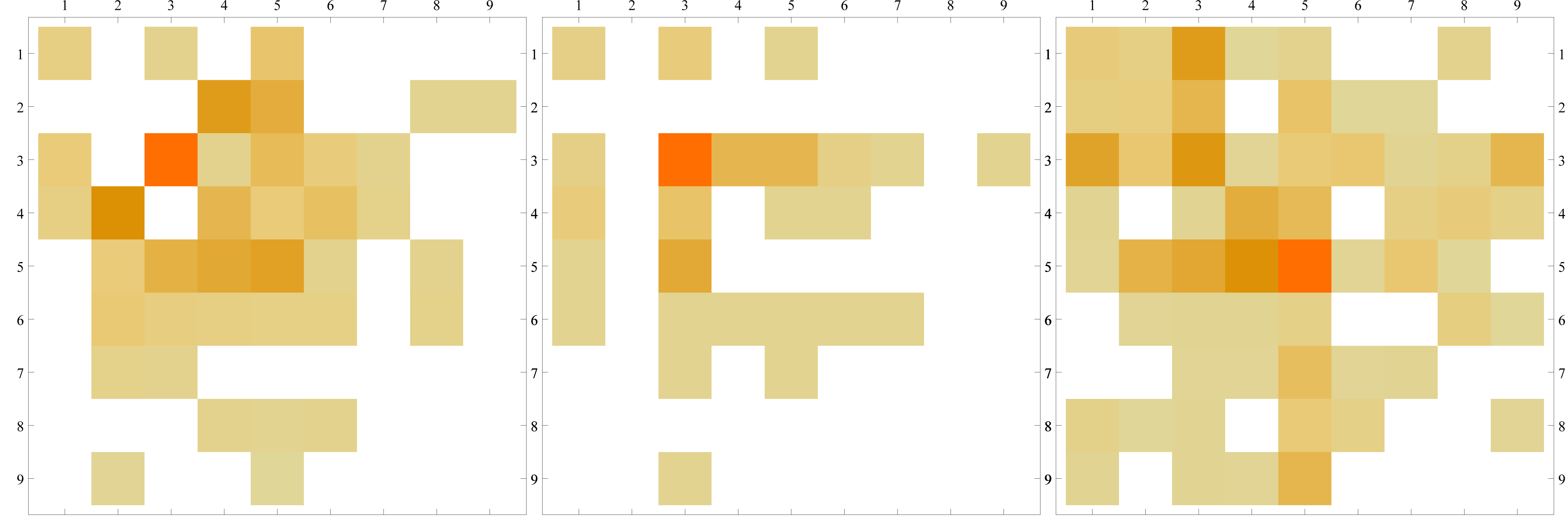}
	\caption{The tempo connection matrices for seeded with three melodies from three separate genres: A is Jazz standards, B is the rock'n'roll and C is light opera.}
	\label{fig:ComboMusicTempo}
\end{figure}

The combined connection matrices for the note connections and note lengths are shown in figures~\ref{fig:ComboMusicNote} and \ref{fig:ComboMusicTempo} respectively. Simple examination of these matrices can tell us a few things about the differences between these musical styles. Looking at the tempo changes in figure~\ref{fig:ComboMusicTempo} we see that the jazz standards make the most use of different length notes with the crochet and quaver being the most popular. Both the rock'n'roll and the light classical are `faster' as they use quavers overwhelmingly (note that this doens't apply to all music in this genre, we chose rather fast classical and Elvis' more dancable tunes). Oddly, the light classics had less variation than rock'n'roll in the timing.

As the graphs shown in figure~\ref{fig:ComboMusicNote} are less easy to understand at a glance. Elvis's rock'n'roll tracks tend to focus on either the lower notes or the higher ones. Both the jazz and the light opera avoids the higher notes. We can look at the reducibility of the connection matrices, which is a measure of the minimum number of connections to represent this music (i.e. pruning the unused connections). For light opera melodies we only need 16, 19 for the rock'n'roll and 20 for the jazz standards (perhaps reflecting that the jazz standards were not the product of a single composer or composition team). The matrices are not symmetrical, but are not far off it and the symmetry can be measured by taking the difference between the transpose and the original matrix. All three styles have a similar symmetry, with the light opera being 83\%  symmetrical, the jazz standards are 83.3\% and the rock'n'roll is the least symmetrical at 85.4\%. 

\begin{figure}
	\centering
		\includegraphics[scale=0.5]{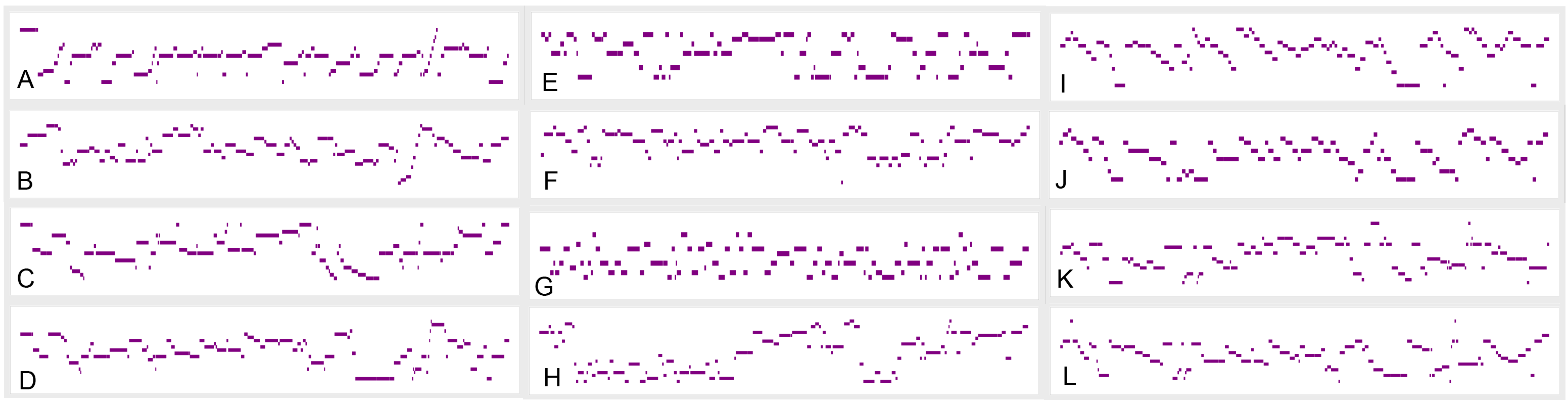}
	\caption{The composed music, output as a result of different seeded connection matrices. A. How high the moon, B. Ain't that a kick in the head, C. The world on a string, D. seeded with all three jazz standards, E, All shook up, F. Burning love, G. Jailhouse rock, H seeded with all three Elvis tunes, I. Modern Major General, J.  When a Felon, K. Better far to live and die, L connection matrix seeded with all three gilbert and Sullivan melodies}
	\label{fig:Output}
\end{figure}

Figure~\ref{fig:Output} shows the output music from memristor networks as seeded by single songs (on the top three rows) and connection matrices seeded by all three input songs. Each of these graphs was generated with the same pseudo-random generated numbers. 

Note, that the system we have set up has been only used to generated the melody, however the system as set-up can allow multiple simultaneous connections, i.e. chords. In a real memristor network, we would expect multiple connections to spike at the same time, creating chords.

The separation of tempo from note, allows us to compose music (using both) and change the performance of the piece using the tempo matrix. 

\section{Time variance of this system}

As explained, the time dependence and spiking behaviour will feedback into the structure of the network and thus alter the system over time. This should make a robust and interesting music generating system, but may produce static solutions such as a dead network (no spikes), epilectic network (many spikes in utter synchrony), resting network (spikes in a repetitive, boring, oscillation). As these solutions have not yet been observed in memristor lab tests, we are optimistic about avoiding such solutions. Nonetheless, all three of these stable state solutions can be reset by wiping and reseeding the network. 

\section{Conclusions}

We have demonstrated a novel approach to music generation networks that relies on non von Neumann hardware and is computationally intractable to solve. We have shown with our simplifications that it is worth pursuing. We have identified properties of memristors which are useful for such a network and tested simple 3 memristor proof-of-concept prototypes. 

The problem of building a memristor-based music composer is an interesting one as it involves many of the same challenges as building a memristor-based brain. Perhaps the creation of recognisable music is a good test of the creation of a working brain. Furthermore, the problem of creating music is far more tractable than that of creating a brain, as every human can recognise a good end solution ('Does it sound like music or noise?), whereas no one can really recognise a working intelligence from looking at the spikes in a neuronal network. 

The creation of a 7,744 memristor network is a little beyond the current state of the art. In our lab, memristors are synthesised by the hundred and have to be wired together by hand. However, such a network is not that far off at places like HP where they have successfully synthesied many thousands of memristors in a neural memristor chip. It is currently outside of the price range of a music composer, but we anticipate the price to come down in the future.

\bibliography{UWELitMusic}

\end{document}